\documentclass[12]{article}
\usepackage{fleqn}
\usepackage{graphicx}
\usepackage{amssymb,amsmath}
\begin{document}
\begin{center}
in Proceedings Group29, edited by C. Bai, J.~P. Gazeau and M. L. Ge, {\it Nankai Series in Pure, Applied Mathematics and Theoretical Physics} (World Scientific, 2013). 
\end{center}
\Large
\begin{center}{\bf
QUANTUM STATES ARISING FROM THE PAULI \\
GROUPS, SYMMETRIES AND PARADOXES
}
\end{center}
\vspace*{-.0cm}
\begin{center}
Michel Planat
Institut FEMTO-ST, CNRS\\
32 Avenue de l'Observatoire, F-25044 Besan\c{c}on, France\\
E-mail: michel.planat@femto-st.fr
\end{center}

\begin{abstract}
We investigate multiple qubit Pauli groups and the quantum states/rays arising from their maximal bases. Remarkably, the real rays are carried by a Barnes-Wall lattice $BW_n$ ($n=2^m$). We focus on the smallest subsets of rays allowing a state proof of the Bell-Kochen-Specker theorem (BKS). BKS theorem rules out realistic {\it non-contextual} theories by resorting to impossible assignments of rays among a selected set of maximal orthogonal bases. We investigate the geometrical structure of small BKS-proofs $v-l$ involving $v$ rays and $l$ $2n$-dimensional bases of $n$-qubits. Specifically, we look at the classes of parity proofs $18-9$ with two qubits (A. Cabello, 1996), $36-11$ with three qubits (M. Kernaghan \& A. Peres, 1995) and related classes.  One finds characteristic signatures of the distances among the bases, that carry various symmetries in their graphs.\end{abstract}



\section{Real rays from the multiple qubit Pauli group and Barnes-Wall lattices}\label{BarnesWall}

An $n$-dimensional Euclidean lattice $L$ is a discrete additive subgroup of the real vector space $\mathbb{R}^n$, endowed with the standard Euclidean product, and spanned by a generator matrix $M$ with rows in $\mathbb{R}^n$. The automorphism group $\mbox{Aut}(L)$ is the set of orthogonal matrices $B$ such that under the conjugation action $U=MBM^{-1}$ by the generating matrix $M$, one has (i) $\mbox{det}~ U= \pm 1$ and (ii) $U$ is an integer matrix \cite{con98}. One is specifically interested by the family of Barnes-Wall lattices $BW_n$ ($n=2^m$ and $m>1$) that generalize the root lattices $BW_4\cong D_4 $ and $BW_8 \cong E_8$ with $BW_{16}$ (the densest known lattices) and higher \cite{Plan11a}. Their automorphismgroup group is the so-called real Clifford group\cite{Plan11b,Nebe06}.

Let us study the relationship between the Barnes-Wall lattices and the real rays arising from the $n$-qubit Pauli group $P_n$ \cite{PlanatJPA}. 
The total number of states/rays appearing as eigenstates shared by the maximal commuting sets $mcs$ of operators in $P_m$ is 
$n L$, where $n=2^m$ and the number of maximal commuting sets is $L=\prod_{i=1}^m (1+2^i)$, as shown in column 3 of table \ref{table1}; the corresponding number of real rays is shown in column 4. The rays form maximal orthogonal bases whose orthogonality graph is defined in columns 5 and 6. One gets the following striking observations that 
\begin{itemize}
\item
The number of real rays identifies to the kissing number $k_n=(2^n+2)k_{n-1}$ (with $k_1=1$) of a Barnes-Wall lattice $BW_n$ ($n=2^m$ and $m>1$), that is isomorphic to 
$D_4, E_8, \Lambda_{16},BW_{32}\cdots$. 
\item
The automorphism group of the orthogonality graph attached to the real rays is that of $BW_n$, that is the real Clifford group $C_n^+=2_+^{1+2m}.\Omega_{2m}^+(2)$, of order $2^{m^2+m+1}(2^m-1)\prod_{j=1}^{m-1}(4^j-1)$, with $2_+^{1+2m}$ the extraspecial group of order $2^{2m+1}$ and $\Omega^+_{2m}(2)$ the orthogonal group [the derived subgroup of the general orthogonal group  $O^{\pm}_{2m}(2)$].
\end{itemize} 

One concludes that a natural frame for the real rays arising from the multiple qubit Pauli group $P_m$ is the Barnes-Wall lattice $BW_{2^m}$.

\begin{table}
\label{Cardinalities and symmetries of the real rays arising from the $m$-qubit Pauli group $P_m$.
}
{\begin{tabular}{@{}r r|r||r|r|r|r|@{}}
 $n=2^m$ &  $\#mcs$ &$\#$rays &  $\#$real rays & aut group & $\#$aut group\\
$2$  &   $3$ & $6$ &   $4$& $D_4$ & $8$\\
 $4$&   $15$ & $60$ &   $24$& $\mathbb{Z}_2^4 \rtimes (\mathbb{Z}_3^2 \rtimes D_4)$  & $1152$\\
$8$&   $135$ & $1080$ &   $240$& $\mathbb{Z}_2^6 \rtimes S_8$ & $2580480$\\
$16$&   $2295$ & $36720$ &   $4320$&  $\cong aut(BW_{16})$& $89181388800$\\
$32$&   $75735$ & $2423520$ &   $146880$&  $\cong aut(BW_{32})$& $\approx 4.8 \times 10^{15}$\\
\end{tabular}}
\label{table1}
\end{table}

\section{The Bell-Kochen-Specker theorem for multiple qubits}

The Bell-Kochen-Specker proof demonstrates the impossibility of Einstein's assumption, made in the famous Einstein-Podolsky-Rosen paper \cite{Eins35}
 that quantum mechanical observables represent 'elements of physical reality'. More specifically, the theorem excludes hidden variable theories that require elements of physical reality to be non-contextual (i.e. independent of the measurement arrangement) \cite{Kochen67}.

 A non-bicoloring BKS proof consists of a finite set of rays/vectors that cannot be assigned truth values ($1$ for true, $0$ for false) in such a way that (i) one member of each complete orthonormal
  basis is true and (ii) no two orthogonal (that is, mutually compatible)
 projectors are both true. 

 A parity proof of BKS theorem is a set of $v$ rays that form $l$ bases ($l$ odd) such that each ray occurs an even number of times over these bases. A proof of BKS theorem is ray critical (resp. basis critical) if it cannot be further simplified by deleting even a single ray (resp. a single basis). 
 
The smallest state-independent proofs in three dimensions are of the $31-17$ type ($31$ rays located on $17$ orthogonal triads).\cite{Peres93}. The  smallest BKS proof in dimension $4$ (resp. $8$) is a parity proof and corresponds to arrangements of real states arising from the two-qubit (resp. three-qubit) Pauli group, more specifically as eigenstates of operators forming Mermin's square (resp. Mermin's pentagram) \cite{Cabello96}. 

Apart from the use of standard graph theoretical tools for characterizing the ray/base symmetries, we shall employ a useful signature of the proofs in terms of Bengtsson's distance $D_{ab}$ between two orthonormal bases $a$ and $b$ defined as \cite{Raynal11}
%
$$D_{ab}^2=1-\frac{1}{d-1}\sum_{i,j}^d \left(\left|\left\langle a_i|b_j \right\rangle\right|^2-\frac{1}{d}\right)^2.$$
%
The distance vanishes when the bases are the same and is maximal (equal to unity) when the two bases $a$ and $b$ are mutually unbiased, $\left|\left\langle a_i|b_j \right\rangle\right|^2=1/d$, and only then. We shall see that the bases of a BKS proof employ a selected set of distances which seems to be a universal feature of the proof.

\subsection{Small BKS parity proofs for two qubits from the Mermin's square}

\begin{equation}
\begin{array}{ccc}
| & | & || \\
-Z_1- & Z_2- & ZZ- \\
| & | & || \\
-X_2- & X_1- & XX- \\
| & | & ||\\
-ZX -& XZ- & YY- \\
| & | & || \\
\end{array}
\label{Mermin1}
\end{equation}

The simplification of arguments in favour of a contextual view of quantum measurements started with Peres' work \cite{Peres93} and Mermin's report \cite{Mermin1993}. Observe that in (\ref{Mermin1}), the three operators in each row and each column mutually commute and their product is the identity matrix, except for the right hand side column whose product is minus the identity matrix. There is no way of assigning multiplicative properties to the eigenvalues $\pm 1$ of the nine operators  while still keeping the same multiplicative properties for the operators. 

The next step to be able to see behind the scene, and to reveal the simplest paradoxical/contextual set of rays and bases, was achieved by A. Cabello \cite{Cabello96}. It is a $18-9$ BKS parity proof that can be given a remarkable diagrammatic illustration fitting the structure of a $24$-cell \cite{Aravind2011}. 
More generally, it is already known that there exist four main types of parity proofs arising from $24$ Peres rays \cite{Peres93}, that are of the type $18-9$, $20-11$, $22-13$ and $24-15$. Types $20-11$ and $22-13$ subdivide into two non-isomorphic ones $A$ and $B$ as shown in Table 2, fore more details see \cite{Planat2012a}.
The histogram of distances for various parity proofs $v-l$ obtained from Mermin's square is shown in table 2. The distances involved are $D=\{a_1,a_2,a_3,a_4,a_5\}=\{\frac{1}{\sqrt{3}},\frac{\sqrt{7}}{\sqrt{12}},\frac{\sqrt{2}}{\sqrt{3}},\frac{\sqrt{5}}{\sqrt{6}},1\}$. One can check the expected equality $2\sum a_i=l(l-1)$ in each proof.

\begin{table}
{\begin{tabular}{@{}c c| c| c| c| c| c|@{}}
proof $v-l$ &  \#proofs & $a_1$ & $a_2$ & $a_3$ & $a_4$ &  $a_5$\\
$24-15$ & $16$ & $18$ &   $18$ & $9$ &   $54$&   $6$\\
$22-13A$ &$96$ &  $12$ &   $18$ & $3$ &   $42$&   $3$\\
$22-13B$ &$144$ &  $12$ &   $18$ & $4$ &   $42$&   $2$\\
$20-11A$ & $96$ & $6$ &   $18$ & $0$ &   $30$&   $1$\\
$20-11B$ & $144$ & $6$ &   $18$ & $1$ &   $30$&   $0$\\
$18-9$  & $16$ & $0$  &   $18$ & $0$ &   $18$&   $0$\\
\end{tabular}}
\label{table2}
\end{table}

\begin{equation}
\begin{array}{ccc}
 \left(\begin{array}{cc} 1 &2  \\15 & 16    \end{array}\right)- 1-&\left(\begin{array}{cc} 1 &3  \\17 & 18  \end{array}\right)-3-
&\left(\begin{array}{cc} 2 &3   \\21 & 22   \end{array}\right)- 2\\
|_{16}&|~_{18}&|_{22}\\
\left(\begin{array}{cc} 5 &6   \\14 & 16  \end{array}\right)-5- &\left(\begin{array}{cc} 5 &7  \\18 & 20   \end{array}\right)-7-
&\left(\begin{array}{cc} 6 &7   \\21 & 24    \end{array}\right)-6 \\
|_{14}&|_{20}&|_{24}\\
\left(\begin{array}{cc} 11 &12   \\14 & 15   \end{array}\right)-12- &\left(\begin{array}{cc} 10 &12   \\17 & 20 \end{array}\right)-10-
&\left(\begin{array}{cc} 10 &11   \\21 & 24   \end{array}\right)-11 \\
|_{15}&|_{17}&|_{21}\\
 \end{array}
\label{MagicSquare9}
\end{equation}

The diagram for a $18-9$ proof is a $3 \times 3$ square. The $9$ vertices of the graph are the $9$ bases of the proof, the one-point crossing graph between the bases is the graph (\ref{MagicSquare9}), with $\mbox{aut}=G_{72}=\mathbb{Z}_3^2 \rtimes D_4$.
There are $18$ (distinct) edges that encode the $18$ rays, a selected vertex/base of the graph is encoded by the union of the four edges/rays that are adjacent to it.

As for the distances between the bases, two bases located in the same row (or the same column) have distance $a_2$, while two bases not in the same row (or column) have distance $a_4$, as readily discernible from Table 2.
 Indeed, any  proof of the $18-9$ type has the same diagram as (\ref{MagicSquare9}). Similar diagrams can be drawn to reflect the histogram of distances in proofs of a larger size.

\subsection{Small BKS parity proofs for three qubits from the Mermin's pentagram}

Quantum contextuality of a three-qubit system is also predicted in Mermin's report \cite{Mermin1993} in terms of its famous pentagram. Below we display it in a sligthly different as a magic" rectangle.

\begin{equation}
\begin{array}{cccc}
| & | & | & |\\
Z_1 & Z_1 & X_1 & X_1\\
| & | & | & |\\
Z_2 & X_2 & Z_2 & X_2\\
| & | & | & |\\
Z_3 & X_3 & X_3 & Z_3 \\
| & | & | & |\\
=ZZZ =& ZXX =&XZX=& XXZ =\\
| & | & | & |\\
\end{array}
\label{Mermin2}
\end{equation}

Following \cite{Mermin1993}, (\ref{Mermin2}) is a parity proof of the BKS theorem because mutually commuting operators in the four columns multiply to the identity matrix while operators in the single row multiply to minus the identity matrix. Since each operator appears twice in this reasoning, it is impossible to assign truth values $\pm 1$ to the eigenvalues while keeping the multiplicative properties of the operators.

{The histogram of distances for various parity proofs $v-l$ obtained from Mermin's pentagram is given in Table 3. The finite set of distances involved is
 $D=\{a_1,a_2,a_3\}=\{\frac{\sqrt{3}}{\sqrt{7}},\frac{\sqrt{9}}{\sqrt{14}},\frac{\sqrt{6}}{\sqrt{7}}\}$.

\begin{table}
{\begin{tabular}{@{}r r||r|r|r|r|r|@{}}
 proof $v-l$& \# proofs &$a_1$ &  $a_2$ &$a_3$ \\
$40-15$ & $64$ & $20$  &   $30$ & $55$ \\
$38-13$ & $640$ & $12$  &   $30$ & $36$ \\
$36-11$ & $320$ & $4$  &   $30$ & $21$ \\
\end{tabular}}
\label{table3}
\end{table}

\begin{equation}
\begin{array}{ccccccccc}
&&&&2&&&&\\
6 &-& &7 &- &8& &-&9  ~~==~~1 \\
&&10&&&&11&&\\
&&&&3&&&&\\
&4&&&&&&5&\\
\end{array}
\label{Pentagram}
\end{equation}

As previously, a simple diagram illustrates how distances between the bases are distributed. Let us look at the $36-11$ parity proof. The $11$ bases are displayed
as a pentagram (\ref{Pentagram}) plus the isolated reference base $1$.

 Two adjacent bases of the pentagram have two rays in common. The reference base has with each of the bases on the horizontal line of the pentagram
 four rays in common and is disjoint from any other base. The maximal distance $a_3$ is that between two disjoint bases; the intermediate distance $a_2$
  occurs between two bases located in any line of the pentagram; and the shortest distance $a_1$ is that between the reference base and each of the four bases on the horizontal line of the pentagram.

 \section{Conclusion}
 
 We have performed a systematic investigation of small state parity proofs of the BKS theorem involving real rays of two and three qubits. Small non-parity proofs for four and five qubits are investigated in \cite{Planat2012a,Planat5QB}.  Another ongoing work is the relation of the BKS operator proofs for three qubits and the geometric hyperplanes of the generalized hexagon $G_2(2)$ \cite{SanigaHexagon}. Further work is also necessary to understand the link between the real rays coming from the Pauli group, Barnes-Wall lattices and the BKS state proofs.




\end{document}